\documentstyle{book}
\newcommand{\hide}[1]{}
\begin{document}

\markboth{{\bf From Atoms to Quarks and Beyond: A historical panorama
}} {{G. Rajasekaran}}

\begin{center}
{\large{\bf FROM ATOMS TO QUARKS AND BEYOND:A HISTORICAL PANORAMA}}
\vskip0.5cm
{\bf G. Rajasekaran}
\vskip0.35cm
{\it Institute of Mathematical Sciences, Madras 600113.\\
e-mail: graj@imsc.ernet.in}
\vskip0.35cm
\end{center}

\vspace{1.5cm}
\underline{Abstract} : The inward bound path of discovery unravelling
the mysteries of matter and the forces that hold it together has culminated
at the end of the twentieth century,in a theory of the Fundamental
Forces of Nature based on Nonabelian Gauge Fields, called the Standard
Model of High Energy Physics. In this article we trace the historical
development of the ideas and the experimental discoveries on which
this theory is based. This involves the following components: quantum
field theory and quantum electrodynamics, weak interactions and parity
violation, strong interactions and quarks, nonabelian gauge fields
and spontaneous symmetry breaking. We develop these different strands
of history and weave them together. We also mark significant Indian
contributions wherever possible. Finally we have a glimpse at future
developments, in particular quantum gravity and string theory. An
Appendix on more Indian contributions is added at the end.\\

\vspace{2cm}
\underline {Plan of the article}

\begin{enumerate}
\item  Scope
\item HEP before and after the Standard Model
\item Future of HEP
\item Present Status in India and Suggestions for Future
\item Some Reflections on the Panorama of HEP
\item Appendix: More Indian contributions

\end{enumerate}
\newpage

\section{Scope}

The earlier part of the 20th Century was marked by two revolutions that
rocked the Foundations of Physics.

\begin{center}

\begin{tabular}{|c|} \hline \\

1. \ Quantum Mechanics  \ \& \ 2. \ Relativity \\

\\ \hline

\end{tabular}

\end{center}

\noindent Quantum Mechanics became the basis for understanding Atoms,
and then, coupled with Special Relativity, Quantum Mechanics provided
the framework for understanding the Atomic Nucleus and what lies inside.

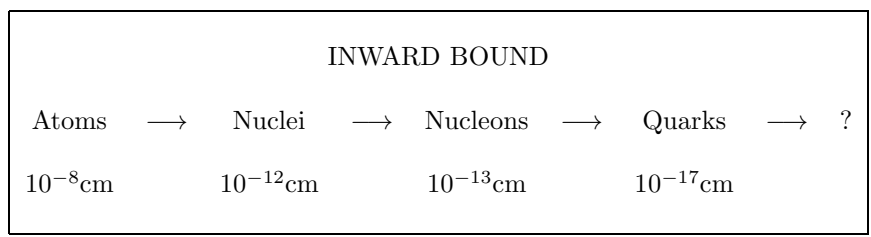
\begin{center}
\begin{tabular}{|ccccccccc|} \hline
& & & & & & & & \\
\multicolumn{9}{|c|}{INWARD BOUND} \\
& & & & & & & & \\
Atoms & $\longrightarrow$ & Nuclei & $\longrightarrow$ & Nucleons &
$\longrightarrow$ & Quarks & $\longrightarrow$ & ? \\
& & & & & & & & \\
10$^{-8}$cm &  & 10$^{-12}$cm & &  10$^{-13}$cm & &  10$^{-17}$cm &  &
\\
& & & & & & & & \\ \hline
\end{tabular}
\end{center}

\noindent
This inward bound path of discovery unraveling the mysteries of matter
and the forces holding it together -- at deeper and ever deeper levels
-- has culminated, at the end of the 20th century, in the theory of {\it
Fundamental Forces} based on {\it Nonabelian Gauge Fields}, for which we
have given a rather prosaic name :

\begin{center}
\begin{tabular}{|c|} \hline \\
THE STANDARD MODEL OF HIGH ENERGY PHYSICS \\
\\ \hline
\end{tabular}
\end{center}

\noindent
But, this is not the end of the road. More on that, later.
Thus,what is called High energy physics (HEP) 
is just the continuation of the era of discoveries that saw
the discovery of the {\it electron}, the discovery of {\it
radioactivity} and {\it X rays}, the discovery of the {\it nucleus} and
the {\it neutron} and the discovery of {\it cosmic rays} and the {\it
positron}.\\

These discoveries went hand in hand with the development of {\it Quantum
Mechanics, Relativity} and {\it Quantum Field Theory}. For, without the
conceptual advances made in these theoretical developments, the above
experimental discoveries could not have been assimilated into the {\it
framework of Physics}.\\

So, the present-day HEP must be regarded as the successor to Nuclear
Physics, which in turn was the successor to Atomic Physics :
\begin{center}

\begin{tabular}{|c|} \hline \\
Atomic Physics $\longrightarrow$ Nuclear Physics $\longrightarrow$ High
Energy Physics \\
\\ \hline
\end{tabular}
\end{center}

\noindent HEP is the front end or cutting edge of the human intellect
advancing into the unknown territory in its inward bound journey.
This {\it 100-year-long history} must be viewed together, to get a true
picture of HEP. It is within this broad framework that we must place any
particular contribution or the totality of Indian contributions, for a
proper perspective. Viewed in this light, it is perfectly natural to
include the great Indian contributions made in the earlier part of the
20th century. {\it If Bose, Raman \& Saha were alive and young today,
they would be doing HEP}. So, I start with their contributions \ldots

\smallskip
\begin{enumerate}
\item M N Saha's (1923) theory of thermal ionization played a crucial
role in the elucidation of stellar spectra and thus was of fundamental
importance for the progress of Astrophysics.
(Saha's (1936) reinterpretation of Dirac's quantization condition for
monopoles, in terms of angular momentum quantization, was very original and its
importance is now recognized.)

\smallskip
\item S N Bose (1924) discovered Quantum Statistics even before the
discovery of Quantum Mechanics by Heisenberg \& Schr\"{o}dinger one year
later. Logically pursued, Bose's discovery by itself would have led to
Quantum Mechanics. But, History went differently. QM was discovered soon
(in fact, too soon) and the flood gates were open. This was unfortunate
for India. This may be called {\bf Missed Opportunity I.}

\smallskip
\item C V Raman (1928) discovered the inelastic scattering of photon on
bound electrons and thus took the concept of photon one step higher.
Raman effect is a fundamental experimental discovery that has not been
surpassed or even equalled in its importance and impact even after 70
years by any other experiment done in this country.

\smallskip
\item S.Chandrasekhar (1932) applied relativistic quantum mechanics to
the interior of stars. He calculated the degeneracy pressure (or Pauli
pressure) of a relativistic electron gas and thus initiated our
understanding of the gravitational collapse of stars.

\smallskip
\item H.J. Bhabha's (1935) calculation of $e^+e^-$ scattering was one of
the earliest nontrivial applications of Dirac equation to a process in
which Dirac's hole theory played a crucial role. This was done even
before a full-fledged Quantum Field Theory existed.
\end{enumerate}

\newpage

\section{HEP before and after the Standard Model}
\subsection{QED}
\begin{center}
\begin{tabular}{|c|} \hline \\
$L \ = \ - \frac{1}{4} (\partial_\mu A_\nu - \partial_\nu A_\mu)^2 \ -
\ \bar{\psi} \gamma^\mu (i \partial_{\mu} - e A_\mu) \psi - m \bar{\psi} \psi $
\\
\\ \hline
\end{tabular}
\end{center}

\begin{tabular}{lcl}
Planck (1900) & : & Quantization of radiation energy \\
[3mm]
Einstein (1905) & : & Photon \\
[3mm]
{\it Bose} (1924) & : & Photons as identical particles \\
[3mm]

$\left.\!\!\!\! \begin{tabular}{ll}{\it Bose} and Einstein & (1926)\\
 Fermi and Dirac & (1926) \end{tabular}\right\} $
 & : & Quantum Statistics\\[5mm]

Dirac (1927) & : & lays the {\it foundations for QED} by \\
& & introduction of $a,a^+$ for photons \\
& & \hfill $[a,a^+]=1$ (bosons) \\
[3mm]
Jordan \& Wigner (1928) & : &  $\{b,b^+\} \ = \ 1$ (fermions) \\
[3mm] $ \left.\!\!\!\! {\begin{tabular}{ll} {\rm Compton} &
(1925) \\
{\it Raman} & (1928)
\end{tabular}} \right\}$
& : & photon \ scatters \ like \ particle  \\
[3mm]
Dirac (1928) & : & Relativistic Eq. for electron \\
[3mm]
Anderson (1932) & : & Discovery of $e^+$ \\
[3mm]
{\it Bhabha} (1935) & : & $e^+ e^-$ scattering \\
[3mm]
Kramers (1947) & : & The idea of renormalization \\
[3mm]
Lamb \& Retherford (1947) & : & Exptl discovery of Lamb shift \\
Bethe (1947) & : & First calculation of Lamb shift \\
& & using renormalization
\end{tabular}

\begin{tabular}{llll}
$\left.\!\!\!\!\begin{tabular}{l} Feynman, \\ Schwinger, \\
Tomonoga, \\ Dyson \end{tabular} \right\}$ \  \  & $(1946-50)$ &:&
$\!\!\!\!\left\{ \begin{tabular}{l}Covariant Formalism, \\ Perturbation Series for
S matrix, \\ Feynman Diagrams, \\ QED emerges as a  \\
renormalizable QFT. \end{tabular} \right.$
\end{tabular}

The development of Quantum Electrodynamics (QED) is concurrent with
the development of Quantum Field Theory (QFT),which has been the basic
language of HEP, at least so far.\\

QED is characterised by the Lagrangian density given in the box above
and the various milestones in its development are tabulated below that.
We now describe them in more detail.\\

The history of QED originates with the concept of the photon. Following
Planck's epochal discovery in 1900 that the experimentally observed
black-body radiation required quantization of radiation energy, Einstein
introduced the concept of photon in 1905. But it took many years and the
formulation of quantum mechanics itself in 1923-24 by Heisenberg and
Schrodinger before Dirac could lay the foundation of QED and QFT in
1927, by introducing the annihilation and creation operators a  and
a   for photons satisfying commutation relations. The work of Bose
and Einstein as well as that of Fermi and Dirac led to the recognition
that there are are two fundamentally different types of quantization
and correspondingly particles or quanta come in two varieties,namely
bosons and fermions. It was Dirac who clarifed the situation and
thus unified both kinds of quantum statistics. All this was in 1926.
In 1928, Jordan and Wigner introduced the annihilation and creation
operators for fermions such as electrons satisfying anticommutation
relations, thus completing the picture.\\

Compton's (1925) and Raman's (1928) discoveries that photon scatters
like a particle both from free and bound electrons took the concept
of the photon a step further.\\

The next milestones were Dirac's discovery of the relativistic
wave equation for the electron in 1928 and Carl Anderson's discovery
of the positron in 1932. The concept of the antiparticle emerged
as a law of nature that was a neccessary consequence of relativistic
quantum mechanics. Bhabha (1935) showed  that the scattering of electron
and positron could be calculated correctly and systematically within
the framework of Dirac equation.\\

Modern QED starts with renormalization. The idea of renormalization
was due to Kramers, but the breakthrough came from the experimental
discovery of Lamb shift (1947) which provided the impetus for
the first calculation of this effect by Bethe (1947) using the
idea of renormalization. Through the seminal work of Feynman,
Schwinger,Tomonoga and Dyson, QED emerged as a covariant, local
and renormalizable QFT, capable of precise calculations within
the perturbative framework. Precision experiments on Lamb shifts
and g-2 of electron and muon soon proved QED to be one of the 
most successful and precise theories ever constructed.\\

Here one must refer to S N Gupta's contribution (1950). It was
Gupta who first constructed a manifestly relativistic formulation
of QED. Before him, QED was formulated in Coulomb gauge which
lacked manifest relativistic invariance. Gupta showed how to do
QED in the covariant Lorentz gauge in a consistent way by using
indefinite metric.\\

It may be worthwhile to pause and look back at the earlier
decades before QED was proved to be a correct theory. Both
because of the intrinsic divergence disease of the theory
and of the supposed incapability of the theory to explain
the higher energy phenomena in the realm of cosmic rays,
it was widely expected that QED will fail. Heitler's
well-known book on "Quantum Theory of Radiation" provides
good evidence for this part of history. Succeeding editions
of the book estimated the demise of QED at higher and higher
energies. That demise never came. QED is still alive and kicking
although incorporated in a bigger framework, namely the Standard
Model of High Energy Physics. There is perhaps a moral in this
story--the widely prophesied demise of the Standard Model also
might never come!

\vspace{.5cm}

\subsection{Weak Interactions}

\begin{tabular}{lcl}
Becquerel (1896) & : & Radioactivity $(\alpha, \beta, \gamma)$\\
\\
Pauli (1930) & : & "Neutrino" \\
\\
Fermi (1934) & : & Theory of $\beta$ - decay : $\displaystyle L_{int} \ = \ \frac{G_F}{\sqrt{2}} \bar{p} \gamma_{\mu} n \ \bar{e} \gamma^{\mu} \nu + h.c.$
\end{tabular}\\

\noindent
\begin{tabular}{lll} 
$\!\!\!\! \begin{tabular}{ll}
\  Lee, Yang, Wu  & (1956)\ \ \ \ : \\
{\it Sudarshan} \& Marshak &(1957)\ \ \ \ : \end{tabular}$ & &
$\left.\!\!\!\!
\begin{tabular}{l}
Parity Revolution\\V-A Form 
\end{tabular}   \right\}\gamma_{\mu} \rightarrow \gamma_{\mu} (1-\gamma_5) $ 
\\[5mm]
Feynman \& Gell-Mann  (1957)  \ \ \  \hspace{0.6mm}  : &&  Universal current $\times$ current theory
\end{tabular}

\vspace{5mm}
The story of weak interactions starts with Henri Becquerel's
discovery of radioactivity in 1896 and its subsequent classification
into alpha, beta and gamma decays of the nucleus.But the real
understanding of beta-decay in the sense we know it now came
only after Enrico Fermi invented a physical mechanism for the
beta-decay process.\\

The basic ingredient for Fermi's theory had been provided by
Wolfgang Pauli.To solve the puzzle of the continuous energy
spectrum of the electrons emitted in the beta-decay of the nuclei,
Pauli had suggested that along with the electron, an almost
massless particle also was emitted. Fermi succeeded in incorporating
Pauli's suggestion and thus was born the theory of weak interactions.
Fermi also named Pauli's particle as neutrino.\\

Drawing an analogy with QED where the basic interaction is the
emission of a photon by an electron, Fermi pictured the weak
interaction responsible for the beta-decay of the neutron as the
emission of an electron-neutrino pair, the neutron
converting itself into a proton in the process.\\ 

This theory of
weak interactions proposed by Fermi purely on an intuitive basis
in 1934 stood the ground for almost 40 years until it was replaced
by Standard Model. However an important amendment to Fermi's theory
came in 1956. This was the discovery of parity violation in weak
interactions by Lee,Yang and Wu. But Fermi theory survived even
this fundamental revolution and the only modification was to replace
the vector interaction by an equal mixture of vector(V) and
axial vector (A) interaction.This is the V-A form proposed by
Sudarshan and Marshak and others.\\

During 1947-55, many new particles such as muons, pions, kaons
and hyperons were discovered and all of them were found to decay by 
weak interactions. In fact parity revolution itself was triggered
by the famous tau-theta puzzle in the decays of the kaons which
was the culmination of the masterly phase-space plot analysis
of the three-pion decay mode of the kaon by Dalitz. The field of
weak interactions
thus got enriched by a multitude of phenomena, of which nuclear
beta-decay is just one. Weak interaction is indeed a universal
property of all fundamental particles.\\

Remarkably enough, all the weak phenomena, namely the weak decays of 
all the particles could be incorporated in a straight-forward
generalization of the original Fermi interaction. This was
achieved by Feynman and Gell-Mann (1957) in the form of the current
x current interaction: 

\begin {center}
\begin{tabular}{|lrcl|}
\hline &&&\\
\hspace{2cm}&$\displaystyle L_{int}$ & =& $\displaystyle \frac{G_F}{2\sqrt{2}} (J^+_\mu J^-_\mu + J^-_\mu J^+_\mu)$ \\ 
&$\displaystyle J^+_\mu$ & = &$ \displaystyle \frac{1}{2} \bar{u} \gamma_{\mu} (1-\gamma_5)d + 
           \frac{1}{2} \bar{\nu} \gamma_\mu (1-\gamma_5) e + \ldots $\\
&$\displaystyle J^-_\mu$ & = &$ \displaystyle \frac{1}{2} \bar{d} \gamma_\mu (1-\gamma_5)u + 
                    \frac{1}{2} \bar{e} \gamma_\mu (1-r_5) \nu + \ldots$\\
&&& \\ \hline		    
\end{tabular}\end{center}

\vspace{.5cm}
A fundamental experimental discovery - the discovery of CP violation
was made by Fitch and Cronin in 1964, in the weak decays of
neutral kaons. This asymmetry in the basic laws of nature and
its ramifications in Cosmology are current topics of research.

The story of weak interactions is not complete without due 
recognition of the neutrino, especially because of more recent
developments to be described later.\\

Pauli proposed the neutrino in 1930. Although because of the 
success of Fermi's theory based on neutrino emission in
explaining quantitatively all the experimental data on nuclear
beta decays, there was hardly any doubt (at least in theorists'
minds) that neutrinos existed, a direct detection of the neutrino
came only in 1956. This achievement was due to Cowan and Reines who
succeeded in detecting the antineutrinos produced from fission
fragments in nuclear reactors.\\

Subsequently, it became possible to detect the neutrinos from 
the decays of pions and kaons produced in high-energy 
accelerators. It is by using the accelerator-produced neutrinos
that the important experiment proving $\nu_{\mu}$ not to be the same as $\nu_e$ 
was done.

Further, even neutrinos produced by cosmic rays
were detected. The underground laboratory (of TIFR) at the deep 
mines of the Kolar Gold Fields was one of the first to detect
cosmic-ray produced neutrinos. This was in 1965.

\subsection {Strong Interactions}

\begin{tabular}{l}
Yukawa (1934)\\ [2mm]
Heisenberg:, Isospin Symmetry \\[2mm]
Powel, Occialini \ldots: Discovery of $\pi$ (1947) \\[2mm]
$\triangle$ resonance: Fermi (1952)  \\[2mm]
Chew-Low Theory : (1954) \\[2mm]
Discovery of Strangeness: Gell-Mann \& Nishijima (1955) \\[2mm]
Resonances (1957-65) \\[2mm]
S-matrix Theory (1957-62)\ :  G.F.Chew\\ [2mm]
$SU(3):  \left\{ \begin{tabular}{l} Sakata, Gell-Mann, Neeman (1961) \\[2mm]
Discovery of $\Omega^-$ (1964) \end{tabular} \right.$ \\[5mm]
Quarks: Gell-Mann, Zweig (1964) \\[2mm]
Current algebra, PCAC, Chiral symmetry (1965-70) \\[2mm]
Scaling in DIS and partons :  Bjorken, Feynman (1967)\\[2mm]
SLAC expts : Taylor, Friedman, Kendal (1967) \\[2mm]
Discovery of Asymptotic Freedom of NAGT \& Birth of QCD (1973)
\end{tabular}
\baselineskip12pt

\smallskip
\parindent6mm

Strong interactions proved a harder nut to crack.After the discovery of
the neutron and the recognition of the strong nuclear force of range
10 Fermi, Yukawa (1934) propounded his famous theory of exchange
of a meson of mass about 100 MeV as the mechanism of this short-range
strong interaction. Thus Dirac's quantization of the electromagnetic
field in 1927 which was the birth of QFT, inspired the creation
of the QFT of weak interactions by Fermi and strong interactions
by Yukawa, within the space of another 7 years.\\

However there was a period of confusion. The disentanglement of
the strongly interacting pion (the Yukawa meson) from the weakly
interacting muon took many years and it was finally resolved only
in 1947 when Powel and Occialini unambiguously discovered the
pi-mu decay chain. The discovery of muon is of great importance
since the modern puzzle of the "generations" starts from the muon.\\

The next important milestone was the invention of the notion of
"isospin" by Heisenberg in 1935 (?), motivated by the equality 
of the pp, pn and nn forces. This is the forerunner of all the
internal symmetries that came later and that play such a fundamental
role in present-day Standard Model.  
Although Yukawa made a successful prediction of the pion on the basis
of a QFT for strong interaction, further progress was stalled
because the interaction was too strong to be treated by perturbative
methods. The impetus provided by the nucleon-pion resonance
Delta (1238) (discovered by Fermi and Herbert Anderson in 1954) led to the partially
successful Chew-Low theory that helped to understand the Delta
resonance. Historically this theory was important since it was
the forerunner of the bootstrap model for hadrons. The 60's were
the golden age of hadrons. Hundreds of them were discovered and
under the influence of this deluge of particles, QFT was declared
dead and an alternate philosophy called S-matrix theory was
proposed, its chief proponent being G F Chew. Many important
ideas were developed under its banner: dispersion relations, Regge
poles, bootstrap, nuclear democracy etc. Ultimately this proved
to be a dead end. And a different line of attack spearheaded by
Gell-Mann proved more successful. Starting with SU(3), this led
to current algebra, and then to quarks, which finally led, via
scaling in deep inelastic lepton-hadron collisions and asymptotic 
freedom to Quantum Chromodynamics (QCD). So, back to QFT even
for strong interactions. However one must not conclude the S-matrix
approach was a complete failure. Although it was a failure for
hadrons, it is this approach that gave birth to String Theory!
More about that later.\\

We now turn to a somewhat more detailed version of the tortuous
history of the discovery of quarks as the constituents of the
hadron.\\

Let us start with Heisenberg's isospin symmetry which is based on 
the SU(2) group. The strange behaviour of the kaons and hyperons
discovered in cosmic rays (strong production rates and weak decay 
rates) in the 50's led to the discovery of a new quantum number
strangeness by Gell-Mann and Nishijima in 1955. This is the
first new hadronic "flavour" to be discovered-- to use the modern parlance.
Sakata and his collaborators invented the SU(3) group symmetry
underlying all the hadrons.This was a consequence of their
speculation that all the hadrons (strange as well as nonstrange)
were composed of three basic hadrons p,n and $\lambda$ (the strange
baryon). A more elegant and highly successful classification
of all the hadrons under Sakata's SU(3) was achieved by Gell-Mann
and Neeman in 1961 when they realised that p,n and $\lambda$
together with 5 other baryons belonged to the octet representation
(The Eightfold Way) of SU(3) rather than the triplet as originally
proposed in the Sakata model. The discovery of $\Omega^-$ in 1964
confirmed the eightfold way and the SU(3) symmetry.\\

But then the following question arose. Eventhough p,n and $\lambda$
were not a triplet, SU(3) group has a triplet, which is the
fundamental representation of SU(3).Where are they? Here one must
describe an event that occurred in Bangalore in August 1961. The
first TIFR Summer School in Theoretical Particle Physics was held
in the Indian Institute of Science Campus at Bangalore. Dalitz
and Gell-Mann were the lecturers.Apart from graduate students,
senior physicists like H J Bhabha, M G K Menon and S N Biswas
were also in the audience. Gell-Mann lectured on SU(3) and the
Eightfold Way, fresh from the anvil, even before they were
published. During one of the lectures, Dalitz questioned him
about the triplets.Why is he ignoring them? Gell-Mann managed
to evade it, inspite of Dalitz's repeated questioning. If
Gell-Mann had answered the question directly, quarks would have
been born in Bangalore in 1961 instead of having to wait for
another three years. If any of the other Indian participants
had succeeded answering, we would have got the quarks and this
would have been a major Indian contribution. This is the
{\bf Missed Opportunity II.}\\

Gell-Mann and Zweig independantly proposed the idea of quarks
in 1964, but it took many years before quarks emerged as physical
entities. Gell-Mann himself was tentative in his proposal; the
title of his published paper was "A schematic model for hadrons".
He took the point of view that quarks are only mathematical.
The three quarks u,d,s replaced p,n,$\lambda$ of Sakata as the
fundamental triplet of SU(3) and all the hadrons are to be regarded
as composites made of quarks and antiquarks.\\

The chief reason for the reluctance to accept quarks as constituents
of hadrons was the prevalent S-matrix philosophy at that time.
The idea of some strongly interacting particles being more
elementary than others was repugnant to the whole scheme of
nuclear democracy. G F Chew in a lecture at The Tata Institute
of Fundamental Research, Bombay in 1967 (?) even claimed to be
able to prove that because of relativity and quantum mechanics
no constituent more elementary than the nucleon and other hadrons
was possible.\\

Inspite of this quark-phobia there were a few bold souls that took
the idea of physical quarks as constituents of hadrons seriously
and worked out the consequences. One may mention the names of
A N Mitra, G Morpurgo and R H Dalitz among others. In particular
Dalitz who was the raporteur for the Berkeley conference in 1965
(?) (one of the famous Rochester series which later became the
the Biennial International Conference in High Energy Physics),
instead of reporting the spin-parity determination for the recently
discovered hadronic resonances, as he traditionally did with
consummate skill and thoroughness, surprised everybody by reporting
how all the spin-parity as well as other quantum numbers of the
hadrons agreed beautifully with the quark model.\\ 

However, although hadronic spectroscopy did give strong evidence
for the correctness of the quark hypothesis and the constituent
quark model, there was no clinching evidence for its correctness.\\

Meanwhile, Gell-Mann's current algebra programme of exploiting
mathematical quarks to abstract the properties of the weak
and electromagnetic currents of the hadrons paid dividends.Coupled
with the notion of PCAC (Partially Conserved Axial Current),current
algebra supplied a temporary theoretical framework for the study
of hadrons.\\

Quark model got a big boost when SU(6) symmetry was discovered
by Gursey,Radicati,Sakita and others. Since each of the three
spin-1/2 quarks has two spin states, there are a total of six
states and SU(6) symmetry follows if one ignores spin-dependant
forces between the quarks. The success of SU(6) brought quarks
nearer to physical reality.\\

But a very serious contradiction developed soon. This is the 
conflict of the apparent total symmetry of the three-quark
wave function in the baryonic ground state with the antisymmetry
requirement of Fermi-Dirac statistics. As a simple example,
consider the doubly charged Delta(1238) which is a 
spin-3/2 baryon made of three u quarks. The wave function
of the three u quarks in the ground state contains a symmetric
spatial part corresponding to zero relative orbital angular
momenta and a symmetric spin part corresponding to total spin 3/2.\\

There was good phenomenological support for the assumption
of symmetric spatial part. SU(6) symmetry assigned both the 
spin-1/2 octet baryons and the spin-3/2 decimet baryons to a
single 56-dimensional representation. Hence Delta (1238) and
the nucleon must have the same spatial part of the quark
wave function. If this spatial wave function were not symmetric,
 it would have nodes corresponding to higher relative orbital
 angular momenta
 for the quarks and such a node in the spatial wave function
 would have been seen in the form factor of the proton. This
 argument developed by A N Mitra and R Majumdar ruled out the
 nodes and hence favoured the symmetrical spatial wave function.\\

 But then the total wave function of the three quarks is symmetric
 under the interchange of space and spin variables, thus violating
 the antisymmetry requirement of the wave function of the three
 identical fermions (uuu). Antisymmetry was restored by the invention
 of a new quantum number, called "colour", which is three-valued,
 and the assignment of an antisymmetric colour wave function
 for the three bound quarks. For, the total wave function is now
 a product of space, spin and colour parts and the total is
 antisymmetric. Greenberg played an important role in the solution
 of this statistics problem, although his original proposal was
 that quarks were parafermions of rank three. It was Nambu who
 suggested the idea of a three-valued quantum number, which was
 later called colour by Gell-Mann.\\

 Quarks became "real" only after two important experimental
 discoveries that came later--scaling in deep inelastic scattering
 (in late 60's) and $\psi$, the bound state of charmed quark $c$ and
 antiquark $\bar{c}$ in 1974.
 Most of the sceptics started believing in the reality of quarks
 only after these developments.\\

 The ultimate triumph of quarks has a parellel in the history
 of atoms. There were sceptics who did not believe in the reality of 
 atoms--Mach, Ostwald and others. Boltzman waged a heroic fight
 against their conservative notions, but the battle was won only after Perrin
 verified Einstein's formula for Brownian motion based on the
 reality of atoms.\\

\subsection{Summary of HEP before the Standard Model (before circa 1971)}

Putting together the ideas of subsections (2.1)-(2.3), we have

\begin{eqnarray*}
L & = & - \frac{1}{4} F_{\mu \nu} F^{\mu \nu} + \bar{e} \left[ i\gamma_\mu
(\partial^\mu - ie A^\mu)-m_e\right] e + i \bar{\nu}_e \gamma_\mu
\partial^\mu \nu_e \\
&&+ \bar{\mu} \left[i \gamma_\lambda(\partial^\lambda - ie
A^\lambda)-m_\mu\right]\mu + i \bar{\nu}_\mu \gamma_\lambda \partial^\lambda
\nu_\mu\\
&& + \bar{u} \left[ i\gamma_\mu(\partial^\mu + \frac{2}{3} i e A^\mu) -
m_u\right]u \ + \ \bar{d} \left[i\gamma_\mu(\partial^\mu - \frac{i}{3}
eA^\mu) - m_d\right]d\\
&& + \bar{s} \left[ir_\mu(\partial^\mu - \frac{1}{3} i e A^\mu) -
m_s\right]s \\
&& + \frac{G_F}{2\sqrt{2}} \left(J_\mu^+ J^-_\mu + J^-_\mu J^+_\mu\right)\\
&& +\mbox{\ strong interactions among quarks whose nature was not known}.
\end{eqnarray*}
where,

$$ J_{\lambda}^{-}={1 \over 2} \bar{e} \gamma_{\lambda}(1 -
\gamma_{5}) \nu_{e} + {1 \over 2} \bar{\mu} \gamma_{\lambda}(1 -
\gamma_{5}) \nu_{\mu}
 + {1 \over 2} (\bar{d} \cos \theta_{c} + \bar{s} \sin \theta_{c})
\gamma_{\lambda} (1 - \gamma_{5}) u $$

$$J_{\lambda}^{+}= (J_{\lambda}^{-})^\dagger $$
and
$$ \sin \theta_{c} \approx 0.22. $$

\noindent
Here we have  the Lagrangian density describing the electromagnetic and weak interactions of
the three quarks $u, d, s$ and the four leptons $e, \mu, \nu_e,\nu_\mu$. The
existence of these quarks as the constituents of the hadrons had been
inferred from hadron spectroscopy through a clever guess. However nobody
knew the form of the strong interaction among the quarks which is
responsible for binding them into hadrons. So, it is left unspecified.
The weak current $ J_\lambda^-$ has been written in term of the Cabibbo-rotated
quarks, in order to incorporate the weak decays of the strange hadrons. CP
violation was experimentally known, but \underline{not} understood theoretically.\\

\subsection{ Brief History of The Standard Model }

\begin{tabular}{llll}
&{\bf Theory} & & {\bf Experiment}\\
1954 & Nonabelian gauge fields &&\\
1964 & Higgs mechanism & & \\
1967 & EW Theory & & \\
& &1968 & Scaling in DIS \\
1971 & Renormalizability of EW Theory \hskip .5cm & & \\
1973 & Asymptotic freedom $~~\rightarrow$ QCD & 1973 & Neutral current \\
& & 1974 & Charm \\
& & 1975 & $\tau$-lepton \\
& & 1977 & Beauty \\
& & 1978 & polarized $e\, d$ expt \\
& & 1979 & 3 jets \\
& & 1983 & W,Z Bosons\\
& & 1994 & Top \\
& & 1998 & $\nu$ mass
\end{tabular}

\vspace{3mm}

The major events which culminated in the construction of the Standard Model
are shown in this table in chronological order. Using nonabelian gauge
theory with Higgs mechanism, the Electroweak (EW) theory was already constructed in 1967,
although it attracted the attention of most theorists only after another 4
years, when it was shown to be renormalizable. The discovery of asymptotic freedom of
nonabelian gauge theory
and the birth of QCD in 1973 were the final inputs that led to the full
standard model.\\

On the experimental side, the discovery of scaling in 
Deep Inelastic Scattering (DIS) which led to the
asymptotically free QCD and the discovery of the neutral current which helped to confirm
the EW theory can be regarded as crucial experiments. To this list, one may add the
polarized electron deuteron experiment which showed that SU(2) x U(1) is the correct group for EW
theory, the discovery of gluonic jets in $ e^{+} e^{-} $ annihilation confirming
QCD and the discovery of W and Z in 1983 that established the EW theory. The
experimental discoveries of charm, $\tau$, beauty and top were also
fundamental for the concrete 3-generation SM, with cancellation of anomalies
and CP violation incorporated (although the last feature was theoretically
discovered by Kobayashi and Maskawa in 1974? itself).\\

However note the blank after 1973 on the theoretical side. Theoretical
physicists have been working even after 1973 and experiments also are
being done. But the tragic fact
is that none of the bright ideas proposed by theorists in the past 30
years has received any experimental support. On the other side, 
experiments  have
only been confirming the theoretical structure completed in 1973.
None of the experiments done since 1975 has made any independent 
discovery (except the discovery of neutrino mass).  If this
continues for long, it will be too bad for the future of HEP. I shall come
back to this point later.

\subsection{ Brief Physics Behind the History }

First about the electroweak sector. The idea of using the beautiful
Yang-Mills nonabelian gauge field theory to construct weak interactions
is a old one, but it was Glashow who identified SU(2)xU(1) as the correct
gauge group for an eletroweak theory. However the stumbling block was
the masslessness of the gauge quanta which would contradict the short
range of weak interactions. So the gauge symmetry had to be broken.
Although the idea of spontaneous breakdown of symmetry (SBS) was
around, Goldstone's theorem which predicted the existence of a 
massless scalar boson (the Nambu-Goldstone boson) as the consequence
of SBS prevented the application of SBS to construct any physically
relevant model. Thus,apparently one had to choose between the devil
(massless gauge boson) and the deep sea (massless scalar boson). It
was Higgs who showed that this is not correct; there is no Goldstone
theorem if the symmetry that is spontaneously broken is a gauge symmetry.
The devil drinks up the deep sea and becomes a regular massive gauge boson.
This is called Higgs mechanism. By combining SU(2)xU(1) Yang-Mills
gauge theory with Higgs mechanism Weinberg and Salam independantly
constructed the successful electroweak theory.

At this point one could ask why SBS ? Why not break the gauge symmetry
explicitly? The answer is renormalizability. Although the original
Yang-Mills theory with massless gauge fields is renormalizable,
Yang-Mills theory with massive gauge bosons and with explicit breaking
of the symmetry is nonrenormalizable. Gerard t'Hooft and Veltman
proved that the renormalizability of the original Y-M theory is not
lost if the gauge bosons acquire masses through SBS. 

There was one further problem - the problem of chiral anomalies
originally discovered by Adler, Bell and Jackiw before the
advent of the standard model. These anomalies would make the
electroweak theory norenormalizable unless they are cancelled suitably.
This is what happens; the leptonic and the quark anomalies cancel
each other.

Fermi's theory of weak interactions was not renormalizable and construction
of a renormalizable theory of weak interactions had remained as one of the 
fundamental problems in HEP. The SU(2)xU(1) gauge theory with SBS
solved that problem.

The electroweak theory based on SU(2)xU(1) gauge group links weak and
electromagnetic interactions together. Here a remark on this unification
is in order. Apart from any aesthetic or other considerations, there is
an important physical reason for linking weak and electromagnetic 
interactions. It is possible to construct a renormalizable theory of
weak interactions alone. But since such a theory would neccessarily
contain charged vector bosons (as a part of the gauge boson multiplet),
one has to have a meaningful theory for these electrically charged
vector bosons. It has been known for a long time that the theory of
the charged vector bosons (spin 1 particles) is afflicted with many
diseases. All these prblems are neatly solved, once the charged vector
bosons and the photon (along with another neutral massive vector
boson in the case of SU(2)xU(1)) are combined into the gauge multiplet.
It is the gauge symmetry that cures the diseases.

In electroweak theory, Maxwell's laws of electromagnetism have been
incorporated into a more general system of laws which unify electrodynamics
with weak interactions. The implications of this fundamental unification
are yet to be fully understood or realized. There is no doubt that the
consequences of this unification will be as profound and far-reaching
as Faraday's unification of electricity and magnetism and of Maxwell's
unification of electrodynamics with optics.

Let us next go to QCD. Why QCD ? What is the reason for believing QCD
to be the theory of strong interactions?

As already mentioned in Sec 2.3, for a time physicists had given up
QFT as a useful approach for understanding strong interactions and
taken to the S-matrix approach. So what caused the resurgence of QFT
in strong interaction physics and what is the reason for going for
this particular nonabelian gauge field theory based on SU(3), which
is QCD ?

The reason came from an experiment, the so-called deep inelastic
scattering (DIS) of electrons on the nucleon, in which the virtual
photon of large momentum-transfer-squared $q^2$ emitted by the electron
probes the structure of the nucleon. It was found that, as observed
by the high $q^2$ probe, the nucleon behaves as if it were composed
of free, point-like constituents (called partons by Feynman). The
electron scatters off each parton, elastically and incoherently.
The incoherent sum of all parton crosssections gave a very good description
of the experimental results. It was a remarkable discovery that
such a complicated process in which many hadrons are produced
in the final state could be described in such a simple fashion.
This discovery was originally made at SLAC with electron scattering
and later similar results were found also for the DIS of neutrinos
on nucleon at CERN.

This phenomenon has a rather close resemblance to Rutherford's famous
alpha particle scattering experiments which led to the discovery
of the nucleus inside the atom. Thomson's spread-out atomic model
would lead to soft scattering (i.e.small scattering angles) only.
Experimentally, Rutherford and collaborators found hard scattering
(large scattering angles), thus showing the presence of "point-nucleus"
inside the atom. In the same way, in DIS, even for large $q^2$ (i.e.
large scattering angle), scattering was observed to take place,
in contrast to what would be expected for a spread-out nucleon.
This led to the discovery of point-like constituents deep inside
the nucleon.

More detailed study of the experimental data revealed that these
partons are in fact quarks; they seemed to have the same spins
and charges as expected for quarks.

Attention should now be turned to the adjective 'free'. In addition
to being point-like, the quark-partons behave as if they are free.
If they were interacting, the free-parton model for DIS would not
have worked.

Now the quarks are bound by tremendous attractive forces to make up
the nucleon. So the interaction between quarks should really be
superstrong. And yet, when observed through high $q^2$ probes, this
superstrong interaction weakens to such an extent that the quarks
behave as free particles.

This was a mystery. On the other hand, this provided an important
clue about the nature of the strong interaction itself. One could
now say that any candidate theory of strong interactions should
satisfy this property, namely it should tend to a free field theory
at high $q^2$. Is there any such theory?

In QFT, it is the renormalization group which provides the required
technique to answer this question. By using renormalization group
one can define a momentum-dependent coupling constant, also
called running coupling constant. So what we need is a theory in
which this coupling constant goes to zero for large momenta.
Such a theory is called {\it asymptotically free}, i.e. it tends
to a free field theory for asymptotic momenta.

To cut the long story short, it was soon discovered that none of the
conventional field theories such as $\phi^4$, Yukawa interaction $\bar{\psi}\psi
\phi$ or QED $\bar{\psi}\gamma_{\mu}\psi A_{\mu}$ is asymptotically free. Of all the
renormalizable quantum field theories, only nonabelian gauge theory
was found to possess the unique distinction of being asymptotically free.
The characterictic nonlinearity of Y-M fields, in particular the cubic
vertex is the essential ingredient that makes the theory asymptotically
free.

So, asymptotically free YM theory emerged as the only choice for a 
theory of strong interactions. Since the colour degree of freedom
with three colours was already there (see Sec 2.3), the gauge group
was taken as SU(3) acting on the 3 colours and QCD was born.

Let us next look at the matter sector. All the nucleons are made of
u and d quarks. With electrons to go around the nuclei, all the atoms
can be made. Even the weak interactions are included once the $\nu_e$ is
given. Hence with the quartet made up of a doublet of quarks and a
doublet of leptons $(u,d,\nu_e,e)$, apparently the whole Universe can be made.
But if the Universe started as a fireball as is claimed in the Big Bang
model of cosmology, it would have started with equal 
amount of matter and antimatter,
whereas the present-day Universe seems to be made entirely of matter.
Sakharov speculated that this asymmetry between matter and antimatter
could have been created later during the evolution of the Universe.
One of the neccessary conditions for this is CP violation which certainly
exists in HEP. But does the SM have it? Yes, if the above matter
quartet is repeated atleast twice more. This was the theoretical
discovery made by Kobayashi and Maskawa. CP violation arises from
the complex coupling constants, but often the phases can be absorbed 
into the definition of the fields. 
All the phases cannot be absorbed, if there exist atleast three
quartets. That is why we have, in addition,  
$(c,s,\nu_{\mu},\mu)$ and $(t,b,\nu_{\tau},\tau)$. Together these are now called the three
generations of fermions and all of them have been experimentally
proved to exist.\\

\newpage
\subsection{ The Standard Model of High Energy Physics }

The complete Lagrangian of the Standard Model is given by
\begin{eqnarray*}
 {\cal L} & =& -{1 \over 4}(\partial_{\mu}G_{\nu}^{i} - \partial_{\nu}G_{\mu}^{i} - g_{3}
    f^{ijk}G_{\mu}^{j} G_{\nu}^{k})^2  \\
      &&  -{1 \over 4}(\partial_{\mu}W_{\nu}^{a} - \partial_{\nu}W_{\mu}^{a} - g_{2}   \epsilon^{abc}W_{\mu}^{b} W_{\nu}^{c})^2 - {1 \over 4} (\partial_{\mu} B_{\nu}
         - \partial_{\nu} B_{\mu})^2  \\
	    && -\sum_{n} \bar{q}_{nL} \gamma^\mu (\partial_{\mu} + ig_{3} {\lambda^{i}
	    \over 2} G_{\mu}^{i} +ig_2 {\tau^a \over 2} W_\mu^a +{ig_1 \over 6}
	    B_{\mu}) q_{nL}  \\
	    &&   -\sum_{n} \bar{u}_{nR} \gamma^\mu (\partial_{\mu} + ig_{3} {\lambda^{i}
	    \over 2} G_{\mu}^{i} + i{2 \over 3} g_1 B_\mu) u_{nR}  \\
	    &&   -\sum_{n} \bar{d}_{nR} \gamma^\mu (\partial_{\mu} + ig_{3} {\lambda^{i}
	    \over 2} G_{\mu}^{i} - i{g_1 \over 3} B_\mu) d_{nR}  \\
	    && -\sum_{n} \bar{l}_{nL} \gamma^\mu (\partial_{\mu} + ig_{2} {\tau^{a}
	    \over 2} W_{\mu}^{a} - i{g_1 \over 2} B_\mu) l_{nl}  \\
	    &&   -\sum_{n} \bar{e}_{nR} \gamma^\mu (\partial_{\mu} - ig_{1} B_\mu)
	    e_{nR}  \\
	    && + | (\partial_{\mu} + ig_{2} {\tau^{a} \over 2} W_{\mu}^{a} - i{g_1 \over 2}B_\mu) \phi |^2 - \lambda(\phi^+ \phi - v^2)^2  \\
	    && - \sum_{m,n} (\Gamma_{mn}^u \bar{q}_{mL} \phi^c u_{nR} + \Gamma_{mn}^d
	    \bar{q}_{mL} \phi d_{nR} + \Gamma_{mn}^e \bar{l}_{mL} \phi e_{nR} + h.c.)
	    \\
	    \end{eqnarray*}

The first three terms describe the pure gauge field part of the
SU(3)xSU(2)xU(1) nonabelian gauge theory. The next group of terms
describes the fermions. The left-handed SU(2) doublet quark of
the nth generation (n=1,2,3) is denoted by $q_{nL}$ and the corresponding
right-handed singlets are denoted by $u_{nR}$ and $d_{nR}$. For the leptons,
$l_{nL}$ is the doublet while $e_{nR}$ is the singlet.\\

The last group of terms describes the Higgs field and its 
interactions with itself, with the gauge bosons and with the fermions
which are respectively responsible for spontaneous symmetry breaking,
generation of W and Z masses and generation of masses for the quarks
and leptons. The $\Gamma$'s are complex numbers and hence lead to CP
violation.\\

{\bf Standard Model} is the basis of {\it all} that is known
in HEP. Although it is believed that SM is only an effective
low energy description and it is to be replaced by something
beyond, so far SM has resisted all attempts at overthrowing
it. All the precision tests performed so far are in beautiful
agreement with SM. All the experimental signals that seem
to signal its overthrow, disappear in about 6 months -- 1
year, except one signal, namely the signal that {\bf
neutrino} has mass.
Neutrino is the only particle, a part of which, its right-hand 
part $\nu_R$, has zero quantum number and so it is not acted on by
the SM group : $SU(3) \times SU(2) \times U(1)$. So, $\nu_R$ is banished 
from SM and as a consequence, the mass
of neutrino within SM is zero. That is why $\nu$ having a
mass is regarded as a signal beyond SM.\\

Note the almost complete absence of Indian contribution. (Of course Salam's
name is there, as a major contributor to the construction of the $ SU(2)
\times U(1) $ electroweak theory. We shall eschew parochialism and include
him since he is from the subcontinent.) Let me give  a little bit of my
side of the story here. I was aware of the beauty of Yang-Mills (YM) theory from the
time of Sakurai's famous Annals of Physics paper of 1960 and I realized
the importance of YM theory to weak interaction ever since I listened to
Veltman in the Varenna School in 1964 where he stressed the
conservation of weak currents. When Weinberg's paper with the quaint title
"A model of Leptons" came out in Physical Review Letters in 1967, I was
immediately convinced that this was the correct theory for weak
interactions and began to work on it. I still missed the boat completely
because I was too muddle-headed and stupid. Instead of trying to
renormalize the divergences away (which we now know to be the right thing,
after t'Hooft showed it in 1971), I was trying to generate the strong
interactions from the divergences. I was too ambitious and missed the 
real thing.\\

Mine was a double failure. Since I was very familiar with partons, scaling
and the quark-model sum rules that the DIS structure functions were found
to obey, I was fully aware of the serious problem that was staring at
everybody's face, namely, how to reconcile the free-quark behavior
exhibited by the DIS structure functions with the superstrong interactions
of quarks inside the hadrons. The techniques that were subsequently used to
effect the reconciliation were also known to me. In fact I was giving
a series of lectures on Wilson's RG ideas and the Callan-Symanzik $
\beta $ function at TIFR, when the preprints of Politzer and
Gross-Wilczek proving asymptolic freedom in YM theory came out.\\

So, I failed on both fronts : on both the two most important QFT 
discoveries of the latter part of the $ 20^{th} $ century - namely
renormalizability of YM theory with SSB and asymptotic freedom of YM
theory - both of which being the essential theoretical inputs in the
construction of the SM of HEP. This is {\bf Missed Opportunity III.}\\

Forgetting about myself, it was a collective failure of the Indian High
Energy Physicists. By that tme we had strong theory groups in the country
and we should have made significant contributions in the construction of
SM, but we did not. In my opinon, this is a glaring failure and we cannot
 forgive ourselves for it.\\

The author apologises for the above intrusion of his personal story.
There is no other way of conveying the flavour and excitement of the 
discoveries of the early 70's and there may be lessons to be learnt in
this story of failures.\\ 

\subsection{ Discovery of Neutrino Mass }

Neutrino mass was discovered in 1998 through neutrino oscillations
by the SuperKamioka group in Japan. This discovery was made in the study
of cosmic-ray produced neutrinos, called atmospheric neutrinos.
Inspite of the fact that the Indian group was a pioneer in atmospheric
neutrino physics being the first to detect the atmospheric neutrinos
in the KGF mines in 1965, we lost the initiative and missed the chance
to make this fundamental discovery. This is {\bf Missed Opportunity IV.}

Indications for neutrino oscillations and neutrino mass came first
as early as 1970 from the pioneering solar neutrino experiments
of Davis etal in USA which were later corraborated by many other
solar neutrino experiments. But the clinching evidence that solved
the solar neutrino problem in terms of neutrino oscillations had to
wait until 2002 when the Sudbury Neutrino Observatory could detect
the solar neutrinos through both the neutral-current as well as the
charged-current modes.

\section{Future of HEP}

Standard Model is not the end of the story. There are too many
loopholes in  it. First of all, there are many interesting
questions and unsolved problems within SM :
\begin{itemize}
\item  Higgs and symmetry breaking
\item QCD and Confinement
\item CP and its violation
\item Neutrinos and their masses and mixings
\end{itemize}

The celebrated Higgs mechanism that breaks the SU(2)xU(1)
gauge symmetry and gives masses to all the elementary particles
according to our present understanding of the SM comes with a price.
A massive scalar boson called Higgs boson has to exist but it has not
been seen experimentally inspite of intensive search. Maybe it will
be seen in the LHC (Large Hadron Collider) that is being built at
CERN. Until it is seen, our understanding of the electroweak sector
of the SM will remain incomplete.

In the QCD sector, colour confinement has not yet been proved.
Asymptotic freedom of QCD and its implied weakness of the QCD
coupling for large momenta led to the recognition of QCD as the
theory of strong interactions. But its complement, namely the
strong QCD coupling for small momenta neccessitates nonperturbative
understanding of QCD. That is a tall order, for our understanding
of QFT has not progressed beyond perturbation theory. Remember
all the low energy hadronic physics falls into this regime of
small momentum transfer and hence strong coupling and QCD has
contributed precious little to this physics. Inspite of
enormous work done (lattice QCD and fast computers) our
understanding of the fundamental mechanism of confinement
remains incomplete. One is tempted to repeat the Churchillian
remark: "In no other field has so much effort been wasted
towards so little effect."

Although CP violation exists in the SM with 3 generations of fermions,
its confrontation with experiments is not complete. Such a confrontation
is in progress through experiments in the "B factories". There
is a serious problem with QCD because of the strong CP violation
that it predicts. This problem is not yet solved.

As already mentioned, the discovery of neutrino mass requires
us to go beyond SM, but we do not know the precise direction to take.
An attractive possibility is the "see-saw" mechanism that
naturally explains the smallness of the neutrino mass simultaneously
linking the tiny neutrino mass to superheavy mass scales and
to physics beyond SM. Hence the name: see-saw. This mechanism
requires the neutrino to be a Majorana fermion for which particle
is the same as antiparticle (in contrast to a Dirac fermion). The
only experimental test for the Majorana nature of the neutrino
is the observation of neutrinoless double beta decay for which
no definitive evidence has been established so far.

The solution of these problems sets the agenda of HEP for the
immediate future.

However, the biggest loophole in SM is the omission of
gravitation, the most important force of nature. Hence, it is
now recognized that {\it quantum gravity} (QG) is the next frontier
of HEP, and that {\it the true fundamental scale of physics
is the Planck energy} $10^{19}$ Gev, which is the scale of
QG.

We are now probing the region with energy $\le 10^3 GeV $. One can
see the vastness of the domain one has to cover before QG is
incorporated into physics. In their attempts to probe this
domain of $10^{3} - 10^{19}$ Gev, theoretical physicists have
invented many ideas such as supersymmetry and hidden dimensions and
based on these ideas, they have constructed many beautiful
theories, the best among them being the string theory, which
may turn out to be the correct theory of QG.

\noindent {\bf String Theory}

String theory is the high-energy-physicist's first successful
attempt to construct a relativistic quantum field theory which
is finite even after including gravitation. This was the
reason for all the exitement and fuss when it first hit the
headlines in 1984. For the first time in history we are glimpsing
at a solution to an age-old problem, namely the problem of constructing
a quantum theory of gravity. So far, any attempted theory of gravity
was afflicted with the worst divergence diseases known in quantum
field theory - much worse than the divergences in QED, QCD and
electroweak dynamics all of which were renormalizable divergences.
Quantum gravity is not renormalizable. String theory solves the
problem by a nonlocal generalization of the usual local quantum
field theory based on point particles. A point particle is
replaced by a one-dimensional object known as a string (either
a open or closed string),with a length of the order of the
Planck length (10$^{-33}$ cm).

Here a brief history of string theory is in order. String theory
was born in attempts to construct the hadronic S-matrix. Starting
with Veneziano's discovery (1968)of a formula for the hadronic scattering
amplitude which had both Regge behaviour and crossing symmetry
the S-matrix approach reached its pinnacle in the construction
of the dual model. However, ironically enough, just at that
juncture, Nambu (1969) identified the spectrum of the dual model
with the vibrations of a string and the S-matrix approach was
soon abandoned in favour of a canonical dynamical formalism for
the string.

Remarkable discoveries followed soon: supersymmetry (in two
dimensions) was discovered in the attempt to include fermionic
excitations in the string spectrum (Ramond 1971, Neveu and
Schwarz 1971). Goddard, Goldstone, Rebbi and Thorn discovered
in 1973 that a consistent relativistic quantum dynamics for
the string required the number of space-time dimensions to be
26 and Gliozzi, Scherk and Olive showed that the same for a
supersymmetric string required 10 dimensions. The point-particle
limit of the open string theory was found to yield Yang-Mills
gauge theory while the same limit of the closed string theory
yielded gravitaion (Neveu and Scherk 1972, Yoneya 1974).

In the hadronic context the above discoveries were regarded
either as irrelevant or bizarre (especially the unseen higher
dimensions). Because of these "troubles", the popularity of
the strings waned after about 1974. An additional and 
important reason for the loss of interest in strings as a
description for hadrons was the discovery of QCD which emerged
as the basic theory of hadrons.

Meanwhile, motivated by the discovery of the connection of strings
to gauge and gravitational fields in the point-particle limit,
Scherk and Schwarz in 1974 proposed to liberate string theory
from its original restricted hadronic context and apply the theory
to the whole world and that liberated string theory from all
the "troubles". All they had to do was to change the string
length from 10$^{-13}$ to 10$^{-33}$ cm.

In 1984 Green and Schwarz discovered that chiral and gravitational
anomalies cancel in the open superstring theory in 10 dimensions
if the gauge group is SO(32) or E8XE8. The discovery of this
miraculous cancellation marked the start of the string
revolution.

For local field theories there already exists the beautiful
Kaluza-Klein idea of unification in which both 4-dimensional
gauge forces and 4-dimensional gravity are derived from
higher dimensional (d larger than 4) gravity. The string-
analogue of this phenomenon is the "heterotic string"
discovered by Gross, Harvey, Martinec and Rohm in 1985,
wherein open strings with Yang-Mills internal indices as well as
pure gravity type closed strings can both be derived from
pure gravity-type higher-dimensional closed strings. The
construction of the heterotic string is a high-point of
the ingenuity of theoretical physicists. Whether Nature
utilizes them is yet to be seen.

Including the heterotic and the original open and closed
strings there are now five string theories that are
anomaly-free and consistent. But all of them live in
10 dimensions. How do
we go from 10 to 4 dimensions? By compactification of 6
of the dimensions. But there are thousands of ways of
doing this. Which is the correct way?
Nobody knows for sure.

After the second string revolution of 94-96 initiated by
Edward Witten, Ashoke Sen and others that led to breakthroughs
such as the discovery of duality linking all the consistent
string theories, string theory has become extraordinarily rich.
It has metamorphosed into M theory and it now includes in its
domain not only strings, but also membranes and mulidimensional
branes.

But, Physics is not theory
alone. Even beautiful theories have to be confronted with
experiments and either confirmed or thrown out. Here we
encounter a serious crisis facing HEP. In the next 10-25
years, new accelerator facilities with higher energies such
as the LHC ($\sim 10^4$ Gev) or the Linear Electron Collider
will be built so that the prospects for HEP in the immediate
future appear to be bright. Beyond that period, the
accelerator route seems to be closed because known
acceleration methods cannot take us beyond about $10^5$ GeV.
It is here that one turns to hints of new physics from {\it
Cosmology, Astroparticle Physics and Nonaccelerator particle
physics}. However, these must be regarded as only our first and
preliminary attack on the unknown frontier. These are only
hints. Physicists cannot remain satisfied with hints and
indirect attacks on the superhigh energy frontier.

\noindent {\bf To sum up the situation :}
There are many interesting fundamental theories taking us to
the Planck scale and even beyond, but unless the experimental
barrier is crossed, these will remain only as Metaphysical
Theories.  It follows that,

\begin{itemize}
\item {\it either,  new ideas of acceleration have to be
discovered},  
\item {\it or, there will be an end to HEP by  about 2020
AD.}
\end{itemize}

Some of the ideas being pursued are laser beat-wave method,
plasma wake field accelerator, laser-driven grating linac,
inverse free electron laser, inverse Cerenkov acceleration
etc. What we need, are a hundred crazy ideas. May be, one of
them will work!

By an optimistic extrapolation of the growth of accelerator
technology in the past 60 years, one can show that {\it
$10^{19}$ GeV can be reached before the end of the 21st century}. (See my
Calcutta talks)\footnote{\noindent
 Perspectives in High Energy Physics, Proceedings of the VIII HEP 
 Symposium, Calcutta (1986), p 399;\\
 The Future of HEP, Particle Phenomenology in the 90's, World Scientific (1992), p1
 } But, this is possible only if newer methods and newer
technologies are continuously invented.\\

\noindent {\bf Another Way Out}

In the past three years, another revolutionary idea is
being-tried -- namely to bring down Planck scale from $10^{19}$
GeV to $10^3$ GeV.  This is the so called TeV scale gravity
which uses large (sub-mm) extra dimensions.  (If we cannot go up
to the mountain top why not ask the mountain
top to come down?) One version of this idea which is popular
is due to Randall and {\it Sundrum}.

This is a very interesting field, with a bewildering variety of
worlds that theorists can construct, as a scan of recent hep-net
will show.

Is Nature so kind and considerate to us, that it would have
brought down the Planck scale for our sake? Only Future can tell.

But, if this turns out to be correct, then Quantum Gravity and
String Theory are not some distant theories relevant at $10^{19}$
GeV, but they are immediately relevant at $10^3 - 10^5$ GeV. So,
it becomes even more urgent to understand String Theories and
assimilate them into Physics!

\noindent {\bf Preons}

A brief look at the history of atoms, nucleons and then quarks
would suggest that preons must be the next natural step. There
may exist in Nature a never-ending layered structure.

As we already mentioned, in the hey-days of S-matrix and bootstrap
philosophy in the early 60's, it was even proposed that perhaps
the end of the road was in sight and that no more constituent
structure beyond the hadrons was possible. But the subsequent
development of physics has shown this to be wrong. We now
believe that hadrons are made of quarks. Are quarks, in turn,
made of preons?

Many preonic models have been proposed but none is as yet required
by experimental data. Down to a distance scale of 10$^{-17}$ cm,
quarks and leptons behave like point particles. Nevertheless,
Nature might have already chosen one preonic model and
future experiments might reveal it!

\section { Status of HEP in India and  Suggestions for the Future}

\noindent {\bf Theory :}
There is extensive activity in HEP theory in the country, spread
over TIFR, PRL, IMSc, SINP, IOP, HRI, IISc, Delhi University,
Panjab University, BHU, NEHU, Gauhati University, Hyderabad
University, Cochin Univesity, Viswabharati, Calcutta University,
Jadavpur University, Rajasthan University and a few other
centres. Research is done in almost all the areas in the field
(see Appendix).
Theoretical HEP continues to attract the best students and as a
consequence its future in the country appears bright. However, it
must be mentioned that this important national resource is being
underutilized. Well-trained HEP theorists are ideally suited to
teach any of the basic components of Physics such as QM, Relativity,
QFT, Gravitation and Cosmology, Many Body Theory,  Statistical Mechanics,
and Advanced Mathematical Physics
since all these ingredients go to make up the present-day HEP
Theory. Right now, most of these bright young theoretical
physicists are seeking placement in the Research Institutions.
Ways must be found so that a larger fraction of them can be
absorbed in the Universities. Even if just one of them joins each
of the 200 Universities in the country, there will be a
qualitative improvement in physics teaching throughout the
country. But, this will not happen unless the young theoreticians
gain a broad perspective and train themselves for teaching-cum
research careers. Simultaneously, the electronic communication
facilities linking the Universities among themselves and with the
Research Institutions must improve. This will solve the
frustrating isolation problem which all the University
Departments face.

\noindent{\bf Experiments:}

Many Indian groups from National Laboratories as well as
Universities (TIFR, VECC, IOP, Delhi, Panjab, Jammu and Rajasthan
Universities) have been participating in 3 major international
collaboration expts :

\begin{itemize}
\item $L3$ expt on $e^+e^-$ collisions at LEP (CERN)
\item $D\ zero$ expt on $\bar{p}p$ collisions at the Tevatron
(Fermilab)
\item Heavy-ion collision expts at CERN and BNL 

\end{itemize}

As a result of the above experience, the Indian groups are well
poised to take advantage of the next generation of colliders such
as LHC and Linear Collider. Already the Indian groups have joined the
CMS and ALICE? international collaboration at LHC. 
It is also appropriate to mention
here that Indian engineers and physicists have contributed
towards the construction of LHC itself.

Thus, the only experimental program that is pursued sofar in the
country is the participation of Indian groups in international
accelerator-based experiments. This is inevitable at the present
stage, because of the nature of present-day HEP experiments that
involve accelerators, detectors, experimental groups and
financial resources that are all gigantic in magnitude.

While our participation in international collaborations must
continue with full vigor, at the same time, for a balanced
growth of experimental HEP, we must have in-house activities
also. Construction of an accelerator in India, in a suitable
energy range which may be initially 10-20 GeV and its utilization
for research as well as student-training will provide this
missing link.

In view of the importance of underground laboratories in $\nu$
physics, monopole search, $p$ decay etc, the closure of the deep
mines at KGF is a serious loss. It is planned to revive underground
neutrino experiments in India. A multi-institutional neutrino
collaboration has been formed with the objective of creating
the India-based Neutrino Observatory (INO) at a suitable site.

Finally, as pointed out in the last section, known methods of
acceleration cannot take us beyond tens of TeV. Hence in order to
ensure the continuing vigor of HEP in the 21st century, it is
absolutely essential to discover new principles of acceleration.
Here lies an opportunity that our country should not miss! I have
been repeatedly emphasizing for the past 20 years that
we must form a small group of young people whose mission shall be
to discover new methods of acceleration.\footnote{IPR is already initiating
research in this area and CAT is training young scientists in accelerator 
technology through SERC Schools.}

To sum up, a 4-way program for the future of experimental HEP in
this country is suggested.

\begin{enumerate}
\item A vigorous participation of Indian groups in international
experiments, accelerator-based as well as nonaccelerator-based.
\item Construction of an accelerator in this country.
\item Creation of an underground laboratory
for nonaccelerator particle physics, especially neutrino physics.
\item A programme for the search for new methods of acceleration
that can take HEP beyond the TeV energies.
\end{enumerate}

\section { Some Reflections on the Panorama of HEP }

\noindent {\bf Twists and Turns of History} 

First we reflect on some
of the interesting twists and turns through which HEP has evolved
over the years. There may be important lessons for the future in
the past history.

One of the strange things in the history of HEP
is what happened to internal symmetries such as isospin, strangeness,
SU(3), baryon number and lepton number all of which were beloved quantum
numbers of particle physicists. Since all these played major roles
in the early history of particle physics, one would have thought
these would be the building blocks on which the edifice of HEP
would be erected.

But this is not what happened. Once the SM was in place, all
these old symmetries lost their fundamental significance. The
empirical importance of isospin (SU(2)) and the old flavour SU(3)
is now understood to be merely due to the smallness of the
u,d,s quark masses. The really fundamental internal symmetry
forming the basis of all the HEP interactions through the
gauge principle is the symmetry SU(3)xSU(2)xU(1) acting on an
entirely different space! While SU(3) acts on colour space
which has nothing to do with the old flavour space, SU(2)
still has some connection with the old isospin but acts on
the left handed parts of the quarks u and d and acts on 
heavy quarks as well as on leptons in contrast to the old
isospin.

It is this twist and turn that converted pre-SM particle
physics into post-SM HEP.

Another related twist and turn is the story of the triplet:
the original Sakata triplet gave way to the quark triplet
$(u,d,s)$ and finally to the colour triplet $(q_{1},q_{2},q_{3})$!

The interplay between the history of weak and strong 
interactions is also full of twists and turns. Yang-Mills
theory was originally used by Sakurai to construct the
theory of strong interactions through the old isospin
symmetry and the old hypercharge and baryon number
quantum numbers. But the more successful application
of Y-M theory was in the construction of the electroweak
theory which lay in the future, for that required the
understanding of spontaneous breaking of symmetry (SBS).
And SBS itself was first discovered in the context of 
strong interactions, namely the broken chiral symmetry
with pion as the pseudo Nambu-Goldstone boson.

Chiral symmetry itself originates from the near-masslessness
of the u and d quarks which is the origin of the old
isospin too, as was already mentioned above.

Finally the cycle was completed by constructing the
Y-M theory of strong interactions with colour SU(3)
and without SBS!

Yet another twist and turn, followed by a fantastic 
jump, is the story of the string:
hadronic string (at 10$^{-13}$ cm) gave rise to the
fundamental string (at 10$^{-33}$ cm), with a jump by
20 orders of magnitude.

The story of baryon number B and lepton number L is
even stranger. The conservation of B that is so basic
for the stability of all matter is not an essential part 
of SM at all! Corresponding U(1) is not contained in the
gauge group of SM. Conservation of B turns out to be
an accidental consequence of the particle content of the
SM. For, with the usual particle content of the SM
(gauge bosons, quarks, leptons and the Higgs doublet)
there is no way of writing a B-violating or L-violating
renormalizable interaction. That is the reason why most
of the attempts at going beyond the SM, such as GUT, SUSY
or giving mass to neutrinos, all of which involve adding 
new particles, end up violating B or L or both.

\noindent {\bf A meta-theorem on Higgs}

Scalar bosons have become the theoretician's tool in
building models. Whether one wants to build models going
beyond SM or wants to explain some perceived discrepancy
of experimental data with SM, one creates a new scalar
sector and invokes the Higgs mechanism for SBS. Here is 
a theorem: One can construct a scalar sector to solve
any problem in HEP! Consequently, hundreds
or thousands of models have been constructed in the last
25 years, many with a similar number of scalar bosons. But not
even a single one (even the original one required in the
electroweak theory) has been seen experimentally!
It is a great irony of Nature that, while QFT of the scalars
is the simplest to teach and that is the way it is usually
taught, none has been observed. Is there a fundamental problem 
with elementary scalars? 

\noindent {\bf Is there a Balmer formula?}

In the SM, all the 12 fermion masses are arbitrary parameters
fixed only by experiment. Perhaps one has to extend SM to include 
a theory of generations for understanding the pattern of the  
fermion masses. Enormous amount of theoretical work has been
done to attack this problem, but there is no memorable result.

However,in 1982, Yoshio Koide found a remarkable empirical relation:

$$ m_{e} + m_{\mu} + m_{\tau}={2\over 3}({\sqrt{m_{e}}} + {\sqrt{m_{\mu}}}
+ {\sqrt{m_{\tau}}})^2 $$

which is satisfied to an accuracy of 1 part in 10$^{5}$. There   
does not exist any other relation of comparable accuracy
in all of HEP (except of course the precision results calculated from
QED and Electroweak theory). But, to this day nobody has
succeeded in deriving the Koide relation from any theory.
Is this the much-needed Balmer formula which can serve as
the guide post for discovering the correct theory of 
generations and thus going beyond the SM?

\noindent {\bf A critique of strings}

String theory does not offer solutions to any known problem in
HEP, such as the generation puzzle, the CP problem, the Higgs
problem etc, nor does it give any clue to the various parameters
of the standard model. This may require the correct choice of
the six-dimensional compact manifold. It may be that if we
know the correct six-dimensional potato with all its warts and
holes, the theory will determine all the masses of the quarks
and leptons and all the parameters of the standard model. But
this is a tall order! The discovery of the correct 6-dim manifold
may require more mathematics than what physicists know. It almost
amounts to saying that if you put in all of mathematics, all of
physics will come out.

So we must grant strings have not done anything for HEP (atleast not sofar).
Hence the real motivation for strings at the present stage lies
elsewhere - infact it lies in Quantum Gravity. Gravity must be
incorporated into the rest of physics. It is intolerable to have
one world where gravity is ignored and quantum mechanics reigns
supreme and another world where gravity cannot be ignored but
use of quantum mechanics leads to meaningless divergent results.

Strings are welcome even as a free invention of the human mind,
since it broadens our horizons and allows us to go beyond the
shackles of point particles and local quantum field theories based
on them.Now we know that string theory includes membranes and branes of
higher dimensions too. So one can imagine a relativistic quantum system  
consisting of a mind-boggling variety of interacting extended objects
living in nine space dimensions and one time dimension. Many of these
extended objects are analogous to the solitons in local quantum field theories.

Although M theory now links all the consistent string theories
through duality, there is still no unique string theory as was
once claimed. In effect, there exist an infinite number of theories since
the number of vaccuum states in string theory or M theory is almost infinite. 
But this is not a
catastrophe! For, the number of consistent local quantum field
theories is also infinite. For instance there is an infinite
number of consistent nonabelian gauge theories based on the
infinite number of available Lie groups. That has not prevented
progress in local quantum field theory and its application
to HEP. Experiment and observation allowed us to make a choice
and that is how Standard Model was constructed. That is the way
physics has progressed sofar and there is no reason to expect
anything different in the future.

\noindent {\bf What about cosmology?}

If current ideas in Cosmology and Astrophysics are correct, then early
universe provides us with a HEP laboratory where particle energies
were not limited by any manmade restrictions. So it is believed by
many that all our theories of HEP can be tested by appealing to 
events in the early universe.

There is no doubt that the era of "precision cosmology" has
dawned with the measurement of CMBR anisotropy whose accuracy 
is awe-inspiring. It is claimed that an accurate determination of
the parameters of the evolutionary model of the universe is
possible. We seem to
have come a long way from Landau's dictum: "Astrophysicists are often
wrong but seldom in doubt." However it is rather worrying that
cosmologists are not able to balance both sides of Einstein's equation.
The right-hand-side needs 25 times more energy and matter than
what we see! Such a huge chunk of mysterious "dark energy and dark matter"
seems too much to swallow. If it stays, it will be one of the greatest
discoveries.

In any case, we know of only one universe and the events
presumably occured only once, that too, quite a long time
ago. Modern Science owes its existence to the advent of
repeatable experiments under controlled conditions
whereas History provides only a single sequence of events.
History cannot be a substitute for Science. Cosmology
cannot provide crucial and definitive tests for fundamental
theories of Physics. On the other hand, laws 
of physics inferred from and tested in laboratory experiments
can and must be applied to the study of the history of the universe.
In other words, the only healthy traffic between HEP and cosmology
is a one-way-traffic:
		     HEP $\longrightarrow$ Cosmology

\noindent {\bf New Technologies?}

All modern technology is based on electrodynamics. We now know
that electrodynamics does not stand alone; it is only part of the
unified electroweak dynamics. What are the deeper implications
of the electroweak unification? Will there be a technology based
on electroweak dynamics? Our understanding of QCD is at an even more
primitive stage, because of colour confinement. But chromodynamics
will be mastered and chromodynamic technology also will come.

As compared to electrodynamics, our understanding of SM is at a
very preliminary stage, perhaps comparable to the study of 
electricity by rubbing amber on wool or glass on cat's fur!
Electromagnetism is just one of the 12 forces contained in the SM which
is based on 12 gauge bosons. One can envisage a stage when all the forces of SM will be
released and put to work. That will be SM technology.

\noindent {\bf New Principles of Acceleration?}

We have already stressed the importance of discovering new principles
of acceleration. There is no future for HEP without it. Supergravity,
Higher Dimensions, String Theory and M Theory are likely to fade
away as Metaphysics if there is no direct experimental support.
Planck energy must be attained in the laboratory. After all, there
is no Law of Nature (such as the Second Law of Thermodynamics)
that forbids this. If there is such a Law, one must prove it.  

All accelerators are based on electrodynamics. There will come a day
when accelerators would be based on SM technology.

\noindent {\bf Why do all this?}

What is the aim of the Inward Bound Journey? Why should we 
delve into deeper regions of space-time? Why should we push
the frontier of HEP to higher and higher energies?

The real justification is the expectation that we will
reach the boundaries of validity of our present view of
the physical universe based on our present conceptual
framework of space-time (namely relativity) and our
present understanding of dynamics (namely quantum mechanics).

\newpage
\section { Appendix: More Indian contributions}  

The following Appendix is presented with apologies for any
omissions for which the author takes the blame.

Seminal and well-known contributions by Indian physicists have
already been mentioned in the main story of the historical
panorama.Many more contributions have been relegated to this
Appendix, chiefly because their inclusion there would have affected
the flow and flavour of the main narrative. Also, attention must
be drawn to the negative comments in Sec 2.5 about HEP in the last
30 years which apply to all the world-wide work of the relevant
period including that by Indian physicists mentioned in this Appendix.

\vspace{3mm}

\noindent {\bf Before Standard Model:}

\vspace{3mm}

The importance of Bhabha scattering (the scattering of positron
on electron) in the development of QED was mentioned in Secs
1 and 2. Bhabha's formula is nowadays routinely used to
calibrate the beams at the electron-positron colliders.
Bhabha's analysis of cosmic rays and his identification of
the penetrating component as being due to a "heavy electron"
with mass around 100 times that of the electron was a
brilliant piece of cosmic ray phenomenology. This is the
particle which we now call the "muon".

Bhabha-Heitler theory of cosmic ray showers is well-known.Their
theory of electromagnetic shower development initiated by a 
high-energy electron or photon cascading through bremstrahlung,
pair production and pair annihilation has become standard
text-book stuff.

The discovery of the V-A form of weak interactions by Sudarshan
and Marshak (see Sec 2.2) was a significant milestone in the
history of HEP.

Regge poles were mentioned in Sec 2.3. These are poles of the 
S matrix in the complex angular momentum plane and played an important
role in the S matrix theory of strong interactions. B M Udgaonkar
derived the basic formulae for the high-energy behaviour of hadronic
cross sections in the s-channel that is controled by the Regge poles
in the t-channel. V Singh studied the Regge poles in the exactly
soluble case of Coulomb scattering. Deser, Gilbert and Sudarshan
wrote down a useful integral representation for the three-point
function. S M Roy derived the basic 
integral equations satisfied by pion-pion scattering from analyticity,
crossing symmetry and unitarity. These, known as Roy equations,
are now used in combination with chiral symmetry, to confront with
experimental data.
 
 In the days of flavour SU(3) symmetry, the Gell-mann - Okubo mass
 formula based on the octet property of SU(3) breaking was an important
 ingredient. Based on the same octet assumption, V Gupta and V Singh
 produced a sum rule for the decay widths of the baryon decuplet.
 When SU(3) got enlarged to SU(6) as discussed in Sec 2.3, M A ? Beg
 and V Singh obtained the SU(6) mass formula. One must also mention
the work of Mitra and Ross who used Galilean invariance for
understanding the "enhanced" heavy meson modes (like N-eta) of
hadronic decays.

R H Dalitz and G Rajasekaran showed that a pole in the S matrix
 is in general followed by a retinue of poles in the complex
 energy Riemann sheet. This discovery of "shadow poles" not only 
 removed a serious obstacle to the application of broken symmetry
 to particle physics but also leads to a reformulation of a basic
 tenet of the S matrix theory.

 In the early days of the quark model the possibility of "molecular
 hadrons" namely composites of the $qqq$ and $\bar{q}q$ states was
 envisaged by Rajasekaran who also formulated an empirical test
 for their identification. This topic has become important in
 the current context of exitement over possible existence of 
 pentaquarks, tetraquarks and other multiquark hadronic states.

 Then came current algebra (see Sec 2.3) and again there were many
 Indian contributions. The soft pion relation for K   form factor
 due to  V S Mathur, S Okubo, L K Pandit and others, the pi pi
 mass difference calculation due to Tapas Das, S Okubo and others,
 the pi-N scattering length formula due to A P Balachandran and
 others and the discovery of a fixed pole in the virtual Compton
 scattering on proton due to S R Choudhury, V Gupta, R Rajaraman
 and Rajasekaran are some of these. Early work of T Pradhan on
 what came to be known later as Schwinger term in the current
density algebra must be mentioned.

P K Kabir made substantial contributions to our understanding of CP
violation. His book "The CP Puzzle" became a standard resource
on this topic.

\vspace{3mm}

\noindent {\bf Towards Standard Model:}

\vspace{3mm}

 The confinement of massless Yang-Mills quanta was conjectured
 by Rajasekaran even before the advent of QCD. The very first
 calculation showing the cancellation of divergences in the
 radiative correction to muon decay in SU(2)xU(1) gauge theory
 also was done by him.

 The first model-independant analysis of the neutral current weak
 interaction data was performed by Rajasekaran and K V L Sarma.
 Their equations,subsequently called "Master Equations" by J J Sakurai,
 played a crucial role in pinning down the coupling constants of the
 interaction as was demonstrated by Sakurai and L K Sehgal.

 Are quarks fractionally charged? The remarkable properties of
 broken-colour QCD with integrally-charged quarks were discovered 
 by Rajasekaran and P Roy as well as J C Pati and Abdus Salam.
 Subsequently, S D Rindani, Rajasekaran and many others tested 
 the viability of this non-standard QCD in a variety of "jet"
 experiments. These studies have uncovered one loop-hole after
 another in the experimental tests usually cited in support of
 the standard QCD with fractionally-charged quarks. 

\vspace{3mm}

\noindent {\bf Within Standard Model:}

\vspace{3mm}

V Soni discovered the existence of quasi-stable solitonic 
configurations in electroweak theory, which came to be known as
"sphalerons" and play a crucial role in the electroweak
baryogenesis. A P Balachandran rediscovered the old "Skyrmions"
and brought it within the context of the modern chiral Lagrangians
and this led to Witten's important work on this subject. R Rajaraman
wrote an excellent book on solitons.

Within QCD the problem of colour confinement and the calculation
of hadronic properties remain hard nuts to crack. 
R Anishetty and H Sharatchandra converted QCD into its dual
form which offers fresh insights into the problem. Following
a different lead based on Bethe-Salpeter formalism Anishetty
has been able to make progress into the mesonic sector.
Harindranath etal have attacked the problem using light-cone
approach.

In Sec 3 we had referred to the strong CP problem. In a
critical reanalysis of the problem H Banerjee, D Chatterjee
and P Mitra have questioned whether such a problem exists.

QGP formation in high temperature QCD has been shown in the
lattice version of QCD. R Gavai and Sourendu Gupta have
made substantial contribution to this study.  
Bikash Sinha made important proposals in the search for
signals of QGP formation in heavy-ion collisions.
On the other hand C P Singh foresaw the many obstacles that
interpretation of heavy-ion experiments would face in 
revealing QGP.

\vspace{3mm}

\noindent {\bf Neutrino Physics: }

\vspace{3mm}

A very significant contribution to neutrino physics was made
by R Cowsik in 1972 when he along with McLelland obtained an
upper bound on the sum of neutrino masses using the early
astrophysical data and the big bang model of cosmology.
R N Mohapatra is one of the originators of the idea of
see-saw for the tiny mass of the neutrino. S Pakvasa is
one of the early physicists to recognize the importance
of atmospheric neutrinos for the study of oscillations.
When neutrino oscillations were indicated by many experiments,
one of the complete analyses of both solar and atmospheric
neutrinos within a realistic three-neutrino framework was
made by M V N Murthy, M Narayan, G Rajasekaran and S Uma
Sankar. This group was the first to include the null results
from the CHOOZ reactor within the three-neutrino framework
and thus obtain a bound on the important mixing angle theta13,    
and also show that the solar and atmospheric neutrino
problems were approximately decoupled. This decoupling led
to a considerable simplification of all the subsequent
phenomenological analyses of the solar, atmospheric,
reactor and accelerator neutrinos. A recent up-to-date
analysis is due to S Choubey, S Goswamy, D P Roy,
and Amitava Raychoudhuri.

\vspace{3mm}

\noindent {\bf Beyond Standard Model: }

\vspace{3mm}

What is the next step after the successful establishment of
the Standard Model based on SU(3)xSU(2)xU(1)? Grand Unified
Theories (GUT) unifying quarks and leptons and also unifying
electroweak and QCD through a semisimple Lie
group SU(5), SO(10) or E6 with a single unified 
coupling constant was considered
as a possible next step. But the very first model unifying
quarks and leptons (albeit not based on a semisimple group)
was the SU(4)x SU(2) x SU(2) model proposed by J C Pati and
A Salam and extensively studied by Pati and R N Mohapatra
especially in the context of the left-right symmetric models.
The group-theoretic work of M L Mehta and Pramod Srivastava
proved very useful in the construction of anomaly-free
gauge models.

An important consequence of GUT is baryon number violation
and proton decay, although this has not yet been observed
experimentally. Here one must refer to the theoretical work
of A N Mitra and  Ramanathan who showed that a correct
normalization of the proton's Bethe-Salpeter amplitude
increased its lifetime by a factor of 100! This normalization
was based on quadratic charge conservation a la
DIS (which has a topological similarity to the Feynman
diagram of proton decay), instead of the more usual linear
charge conservation.

The wide disparity between the scale of the electroweak
symmetry breaking in the SM and the scale of the GUT
symmetry breaking, called the problem of hierarchy, can
be solved by invoking supersymmetry. This important discovery
is due to R Kaul and P Majumdar. N D Hari Dass etal proved
a no-go theorem for the de Sitter compactification of
higher dimensional theories.

Superstring theory could be the correct theory incorporating
Quantum Gravity and the Standard Model in one unifying
framework.Ashoke Sen made an important contribution to 
the second string revolution that involved the
discovery of duality symmetries linking all the consistent
string theories. The alternative
nonperturbative approach to Quantum Gravity called Loop 
Quantum Gravity was pioneered by
Abhay Ashtekar.

A large number of Indian physicists have been involved in
the construction of models that go beyond the SM as well as
in deriving the phenomenological consequences of such models
that could provide evidence for them in the upcoming colliders.
Further, there is a strong
Indian group of String Theorists who have made a mark in the
international scene by their outstanding contributions. 
Both these classes of contributions, going respectively by the names of
phenomenology and formal theory, form a substantial part of
present-day HEP activity in India as well as internationally,
but have to await future experiments for their substantiation. 
(This is the blank space alluded to in Sec 2.5.) 
\vspace{3mm}

\noindent {\bf General:}

\vspace{3mm}

Following the footsteps of Dirac whose relativistic wave
equation describes {all} the fermions in the Standard Model,
many high-spin equations were written down in the literature
(including some by Bhabha), but they were all afflicted by
various kinds of diseases. We have already mentioned the
case of charged spin 1 particles whose problems were solved
only by including them into the Yang-Mills gauge multiplet.
The problems of spin 3/2 particles were solved by S D Rindani
and M Sivakumar through Kaluza-Klein theory. Supergravity
also can solve this problem. Thus the higher symmetries of
Yang-Mills, Kaluza-Klein or Supergravity are necessary for
consistent higher-spin interactions. Nature loves higher
symmetries.

Inspite of the fact that parastatistics for quarks was
abandoned with the advent of QCD there exists the possibility
of new forms of statistics for newer degrees of freedom that
may open up when deeper regions of space-time are probed
with higher and higher energy machines. For instance
Strominger has speculated that quantum blackholes
obey "infinite statistics". A K Mishra and G Rajasekaran
discovered an elegant algebra describing a new form of statistics
called "orthostatistics" that combines infinite statistics
with Fermi-Dirac or Bose-Einstein statistics. S Chaturvedi
solved the problem of calculating the partition functions
for parastatistics which remained an open problem for a
long time.

\vspace{3mm}

\noindent {\bf Experiments: }

\vspace{3mm}

In the early days of experimental particle physics M G K Menon
along with others in the Bristol Group discovered many of the
decay modes of K mesons during their studies of the interactions
of cosmic rays in nuclear emulsion. Lokanathan and Steinberger 
determined the branching ratio of pions decaying into electrons
and neutrinos and this played a crucial role in the formulation
of $ V - A $ theory. 
The TIFR cosmic ray group
pioneered many cosmic ray experiments. As already pointed out in Sec 2, 
they were the first to detect atmospheric neutrinos in
the deep mines of Kolar Gold Fields in 1965.

Other outstanding experiments of fundamental significance are
the search for proton decay in the KGF experiment, the search
for axion in the nuclear reactor experiment at Trombay and
the search for the fifth force using torsion balance (by
R Cowsik, N Krishnan and  Unnikrishnan), although none of
these searches going on world-wide have yielded positive
results so far.

Among the major international experiments in which the
Indian groups have participated are the precision tests
at LEP that established the standard model as a 
renormalizable QFT, the discovery of the top quark at
the Tevatron and the Heavy Ion Collision experiments
at CERN and Brookhaven searching for Quark Gluon Plasma.

The first two-in-one experiment BOREX using Boron 12 to
detect solar neutrinos via charged current and neutral
current modes simultaneously and thus establish neutrino
oscillations if they exist was proposed by S Pakvasa
and R Raghavan, but this experiment did not materialise.
Instead, a similar experiment using deuterium in heavy
water proposed by H Chen did materialize and this was
the experiment of Sudbury Neutrino Observatory which
successfully solved the solar neutrino problem.

A unique low-energy solar neutrino experiment BOREXINO
which will focus on the monochromatic Be neutrino
lines was proposed and led by R Raghavan.

\vspace{3mm}

\noindent {\bf Retrospect: }

\vspace{3mm}

Inspite of the above Indian contributions one must admit
the remarkable absence of great Indian contributions
in the recent history of HEP. Why have we come down?
Where are the equivalents of Bose and Raman in the present-
day HEP? There may be sociological reasons for this, but
this is not the place to go into them. 

Is it possible that
India throws up great names only when Physics goes through
revolutionary development as in the beginning of the 20th
century? If so, the next revolution which may come in the
21st century must be eagerly watched! Remember that the
solution of the Quantum Gravity problem and/or the
formulation of String Theory is still incomplete. They
may usher in the next revolution in Physics and may involve
great contributions from India. There are already signs of this,
in the quality of Indian contributions to String Theory and
Quantum Gravity. We shall close with this optimistic remark.

\end{document}